\documentclass[preprint2]{aastex}

\newcommand\lsim{\lesssim}
\newcommand\gsim{\gtrsim}

\begin{document}

\title{Relaxing constraints on dark matter annihilation}
\author{Leonid Chuzhoy}
\bigskip
\affil{Department of Astronomy and Astrophysics, The University of
Chicago, 5640 S. Ellis, Chicago, IL 60637, USA;
chuzhoy@oddjob.uchicago.edu}

\begin{abstract}
The relic abundance of thermal dark matter particles is generally
assumed to be inversely proportional to their annihilation rate,
which is therefore constrained by the present matter density,
$\left<\sigma_{ann}v\right> \sim 10^{-26}\Omega_{\rm dm}^{-1}\; {\rm
cm^3\cdot sec^{-1}}$. Here we point out that much lower values of
$\left<\sigma_{ann}v\right>$ are possible for heavy dark matter
candidates ($m_X\gsim 10$ TeV) that couple to other particle species
through the electroweak force. With heavy dark matter particles
present the early universe may evolve according to the following
scenario. After an early entry into matter-dominated phase, dark
matter particles form self-gravitating microhalos. Collisional
interaction between dark matter particles and the surrounding
radiation field eventually leads to microhalos gravothermal collapse
and annihilation of most dark matter particles. For sufficiently
heavy dark matter candidates ($m_X\gsim 10$ TeV) the universe can
return to radiation-dominated phase before the nucleosynthesis and
thereafter follow the "standard" scenario.
\end{abstract}


\section{Introduction}
While there is a strong evidence for the existence of non-baryonic
dark matter, little is known about its origins.
 In the currently popular scenario, dark matter particles and antiparticles start in thermal equilibrium with radiation.
 When temperature drops below the particle mass, $m_X$, their comoving density begins an exponential fall until the annihilation
 time-scale, $t_{ann}=(n_X\left<\sigma_{ann}v\right>)^{-1}$, becomes larger than the age of the Universe.
 From this point the comoving density of dark matter particles is assumed frozen. For the relic
 abundance of dark matter particles to be consistent with the present value of $\Omega_{\rm dm}$ their annihilation rate at the
 time of freeze-out must equal (Kolb \& Turner 1989)
\begin{equation}
\label{ann} \left<\sigma_{ann}v\right>\sim  3\cdot 10^{-26}\; {\rm
cm^3\cdot sec^{-1}}.
\end{equation}
Unless dark matter is of non-thermal origin (Chung, Kolb \& Riotto
1998), lower values of $\left<\sigma_{ann}v\right>$ would overclose
the universe and therefore can be excluded.

If dark matter particles interact through any known force, during
the decoupling epoch, when $T\sim m_X$, their annihilation
cross-section can not exceed $\sigma_{ann} \lsim \alpha/T^2\sim
\alpha/m_{X}^{2} $. Using this fact, Griest \& Kamionkowski (1990)
obtained from (\ref{ann}) an upper limit on the mass of dark matter
particles
\begin{equation}
\label{mass} m_X\lsim 300 \; {\rm TeV}.
\end{equation}

However, in this paper we point out that dark matter candidates with
higher mass and lower annihilation cross-section are not necessarily
inconsistent with the present value of $\Omega_{\rm dm}$. Unlike the
standard model, which assumes the dark matter comoving density
freezes out soon after the temperature of the Universe drops below
$m_X$, we show that it is possible for it to drop again later, after
dark matter particles form virialized halos.

\section{Dark matter annihilation inside halos}
Unlike the standard model, where the matter becomes dominant only
around $z\sim 10^4$, in a Universe with low
$\left<\sigma_{ann}v\right>$ and high $m_X$ the initial transition
to matter domination happens much earlier. Consequently the growth
of density perturbations and the formation of the first
gravitationally bound halos would be likewise shifted to a much
earlier epoch.

The collapse of dark matter particles into halos gives a strong
boost to the dark matter annihilation rate. Whereas previously the
time-scale for particle annihilation was increasing faster than the
age of the Universe, freeze-out of the local density inside
gravitationally bound halos causes the reversal of this trend. Once
the Hubble time begins to catch up with the annihilation time-scale,
the dark matter comoving density is no longer constant.

Subsequent evolution of the Universe depends on the nature of the
dark matter particles. For purely collisionless dark matter it can
be shown (see Appendix A) that the comoving matter density
eventually approaches an asymptote, $\rho_m(1+z)^{-3}\propto
t^{-1/4}$. This moderate rate of decline turns out to be
insufficient for the Universe to return to radiation dominated phase
before the nucleosynthesis. However, annihilation inside halos can
be greatly sped up if dark matter particles interact with the
surrounding radiation (made up mostly of relativistic quarks and
leptons).

At present weakly interacting massive particles (WIMPs) are the most
popular candidates for the role of dark matter. The collisional
coupling between WIMPs and other species is usually assumed to be
negligible.  At our epoch this assumption is generally correct, as
the Universe temperature is far below the mass scale of $W^\pm$ and
$Z$ bosons which mediate the weak force. However, in the early
universe, when $T\gsim 300$ GeV, the coupling becomes very
efficient. At high temperatures WIMPs scattering cross-section rises
to $\sigma_{elastic} \sim \alpha^2/T^2$, making the ratio between
the relaxation time and the age of the Universe
\begin{equation}
\label{rel0} \frac{\tau_{rel}}{t_{Hubble}} \sim
\frac{m_X+T}{\alpha^2 m_{Pl}},
\end{equation}
where $m_{Pl}$ is the Planck mass. For $m_X\lsim 10^{13}$ TeV and
$0.1\lsim T \lsim 10^{13}$ TeV this ratio drops below unity,
implying that WIMPs kinetic energy is tightly coupled to the
radiation.

Collisional coupling between dark matter and radiation has a
two-fold effect on the formation of self-gravitating dark matter
halos, which is expected to begin after the matter-radiation
equality. On small scales, $T_{virial}<T_\gamma$, the energy
transfer from radiation to dark matter prevents the collapse of the
low-mass halos. On large scales, $T_{virial}\gsim T_\gamma$, the
effect is the opposite. Since self-gravitating objects have negative
heat capacity, the energy loss to the surrounding radiation field
would lead dark matter particles to collapse deeper into the
gravitational well, thereby raising the virial temperature. Thus the
continuous energy loss to the background radiation results in a
gravothermal collapse. \footnote{During the epoch of reionization,
$6\lsim z\lsim 20$, inverse Compton scatterings between electrons
and the CMB photons have a similar effect on baryonic halos.
However, there the gravothermal collapse is generally avoided either
by energy injection from stars and miniquasars or by hydrogen
recombination, which depletes the free electron abundance.}

The time-scale for particle diffusion in self-gravitating objects,
which in our case also sets the time-scale for gravothermal collapse
of larger halos, is
\begin{equation}
\label{diff} \tau_{coll}\sim \frac{\left<\sigma_{el}v\right>}{Gm_x},
\end{equation}
where $\sigma_{el} \sim \alpha^2/T^2$ is the scattering
cross-section for the weakly interacting particles. If, as generally
predicted by the inflation models, the density perturbation spectrum
is scale-invariant, then the velocity dispersion of dark matter
particles in halos with a size above the damping scales should be of
order $v \sim 0.1c$ and the virial temperature $T\sim 0.01m_X$. Thus
 in large halos the gravothermal collapse proceeds on a timescale
\begin{equation}
\label{coll}  \tau_{coll}\sim {\rm 10^3\left(\frac{m_X}{1
TeV}\right)^{-3}\; sec}.
\end{equation}

If halos consist equally of dark matter particles and anti-particles
(i.e. there is no dark matter asymmetry) at some point during the
gravothermal collapse nearly all of them may annihilate into
radiation. Provided the mass of the particle is greater than $\sim
10$ TeV, halos collapse and the resulting conversion of dark matter
to radiation would end before nucleosynthesis and thereafter the
evolution of the Universe can proceed according to the "standard"
scenario.

\section{Discussion}
It follows from our scenario that the existence of stable particles
with high mass and low annihilation cross-section does not violate
the constraints on the present matter density. If several species of
such particles existed in the early universe, each could have been
eliminated in separate periods of matter domination followed by
return to radiation domination.

To satisfy the constraints on abundance of the light elements the
last reversal to radiation domination must have happened prior to
the nucleosynthesis. Nevertheless it is possible for the
transitional matter dominated phase to leave its imprint on the
presently observed universe. Besides leaving a small remnant of
heavy particles, which at present compose the dark matter, the
transitional matter dominated phase might affect later universe by
creating a population of primordial black holes and baryon
asymmetry.

The violation of CP-symmetry in the early universe (Sakharov 1967)
makes it natural to expect a small differential between the coupling
strengths of particles and anti-particles. Inside collapsing halos
the drag force produced on dark matter particles through elastic
scatterings with the escaping radiation may be different from the
force on dark matter anti-particles. As a result, spatial
segregation of dark matter particles and anti-particles would be
produced. Similarly, a differential drag by the infalling dark
matter particles on the outflowing quarks and anti-quarks (or
leptons and anti-leptons) would produce local over/underabundances
of baryons and leptons. During halos gravothermal collapse the
existence of such segregation would make it impossible for the
entire halo to annihilate into radiation. Therefore the collapse of
particles (or anti-particles) dominating at the halo center should
lead to the formation of a black hole.

Since black holes do not conserve baryon numbers (Hawking 1974;
Zeldovich 1976; Carr 1976), formation of a central black hole would
convert a local particle asymmetry into a global one. Depending on
their initial mass, these primordial black holes may have evaporated
long before the present epoch, possibly further amplifying the
matter-antimatter asymmetry in the process (Dolgov 1982).
Alternatively, if their mass was sufficiently large, significant
population of primordial black holes may still be found in our
Universe.

\vspace{1cm} The author thanks E.W. Kolb, M.S. Turner and M. Beltran
for stimulating discussions.

\section*{Appendix A: Annihilation of collisionless particles in virialized halos}

We consider the evolution of matter density field, assuming the
annihilation cross-section scales as $\left<\sigma_{ann}
v\right>\propto v^{-x}$ at non-relativistic velocities.  While it is
common to assume $x=0$, many new particle candidates with $x=1$ have
been proposed (e.g. Hisano et al. 2004; Profumo 2005; Lattanzi \&
Silk 2008). Here we shall assume that $x$ can take any value in the
physically plausible range $0\leq x<2$.

After decoupling from radiation matter density drops to a point when
the annihilation time-scale, $t_{ann}=(n_x \left<\sigma_{ann}
v\right>)^{-1}$, becomes longer than the Hubble time, so that the
comoving density is frozen.  As long as the density field remains
nearly homogeneous with the density and velocity dispersion falling
respectively as $a^{-3}$ and $a^{-2}$ with the increasing scale
factor, $a$, $t_{ann}/t_{Hubble}$ continues to rise. However, this
trend is reversed after matter collapses into halos.  Inside an
isolated gravitationally bound halos both the local density and
velocity dispersion remain roughly constant, freezing the growth of
$t_{ann}$ until it becomes comparable with $t_{Hubble}$. At this
point a halo would start losing mass at a significant rate via
particle annihilation. The subsequent halo evolution of can be
described as following.

The mass loss from particle annihilation and the resulting decrease
of binding energy would cause the halo to expand, so that the
annihilation time-scale tracks the age of the halo. Because the
dynamic time is much shorter than $t_{Hubble}$, the halo always
remains close to dynamical equilibrium. Since the gravitational
binding energy is proportional to $E_g\propto M^2/R$, where $M$ and
$R$ are the total mass and the radius of the halo, changing the mass
and radius by $dM$ and $dR$ respectively, changes $E_g$ by
\begin{equation}
\label{Eg} dE_g=E_g\left(2\frac{dM}{M}-\frac{dR}{R}\right).
\end{equation}
The change of halo kinetic energy, $E_k$, in the annihilation
process depends on the velocity dependence of $\sigma_{ann}$ and the
particle velocity distribution. Assuming Maxwell-Boltzmann
distribution, we find that the average kinetic energy of an
annihilating particle is $(1-x/2)(m_X\left<v^2\right>/2)$, where
$\sqrt{\left<v^2\right>}$ is the local velocity dispersion. Thus
\begin{equation}
\label{Ek} \frac{dE_k}{E_k}=\left(1-\frac{x}{2}\right)\frac{dM}{M}.
\end{equation}
Combining the virial theorem, $E_g=-2E_k$, with equations (\ref{Eg})
and (\ref{Ek}), we find that the radius of the halo increases with
the falling mass as $R\propto M^{-1+x/2}$. Further, by using the
scalings $\rho\propto M/R^3$, $v\propto\sqrt{M/R}$ and $(\rho
v^{-x})\propto t^{-1}$, we find that the total mass of the halo
falls with time as $M\propto t^{\mu}$, $\mu=-4/(16+2x-x^2)$. For the
entire range $0\leq x<2$, $\mu$ is confined to a narrow range
$-0.25\leq x <-0.235$.

Modeling the Universe as an ensemble of isolated halos, we can
follow the global evolution of matter and radiation density fields
\begin{eqnarray}
\frac{d^2 a}{a dt^2}&=&-\frac{4\pi G}{3}(\rho_m+2\rho_\gamma), \\
\frac{d\rho_m}{dt}&=&-\left(3\frac{da}{adt}-\frac{\mu}{t}\right)\rho_m, \\
\frac{d\rho_\gamma}{dt}&=&-4\frac{da}{adt}-\frac{\mu}{t}\rho_m,
\end{eqnarray}
where $\rho_\gamma$ and $\rho_m$ are respectively, the mean
densities of radiation and matter and $a$ is the expansion factor.
Solving the equations, we find the asymptotic solutions
\begin{eqnarray}
a&\propto& t^{(2+\mu)/3}, \\
\rho_m&=&\frac{2+5\mu+2\mu^2}{12\pi Gt^2},\; \rho_\gamma=\frac{-(2+\mu)\mu}{8\pi Gt^2}, \\
\frac{\rho_m}{\rho_\gamma}&=&-\frac{2(1+2\mu)}{3\mu}.
\end{eqnarray}
For $\mu\approx -1/4$  we get $\Omega_m\approx 4/7$ and
$\Omega_\gamma\approx 3/7$.

\end{document}